\documentclass[epj]{webofc}
\usepackage[utf8]{inputenc}
\usepackage{epsfig}
\usepackage{bm}
\usepackage{color}
\usepackage{dcolumn}
\usepackage{tensor}

\usepackage{svg} % For higher res files

\usepackage{amsmath}
\usepackage{amssymb}
\usepackage{amsthm} 
\usepackage{amscd, amsfonts, mathrsfs}
\usepackage{cases}
\usepackage{verbatim} 
\usepackage[varg]{txfonts} % needed for eps fonts
\usepackage{url}
\usepackage{hyperref}
\hypersetup{colorlinks=true,citecolor=blue,urlcolor=blue,linkcolor=blue}
\usepackage{enumitem}
\usepackage{empheq}
\theoremstyle{plain}
\newtheorem{theorem}{Theorem}

\theoremstyle{definition}

\newcommand{\Umb}{\textrm{\textbf{U}}}
\allowdisplaybreaks

\newcommand{\be}{\begin{equation}}
\newcommand{\ee}{\end{equation}}
\newcommand{\bea}{\begin{eqnarray}}
\newcommand{\eea}{\end{eqnarray}}
\newcommand{\bml}{\begin{subequations}}
\newcommand{\eml}{\end{subequations}}

\begin{document}
%%%%%%% TITLE %%%%%%%%%%
\title{Nonlinear causality and strong hyperbolicity of baryon-rich Israel-Stewart hydrodynamics}

%%%%%% Authors names w/ institutions %%%%%%%%%%
\author{\firstname{Ian} \lastname{Cordeiro}\inst{1}\fnsep\thanks{\email{itc2@illinois.edu}} 
\and 
\firstname{F\'abio S.} \lastname{Bemfica}\inst{2,3}\fnsep\thanks{\email{fabio.bemfica@ufrn.br}} 
\and 
\firstname{Enrico} \lastname{Speranza}\inst{4}\fnsep\thanks{\email{enrico.speranza@unifi.it}} 
\and 
%
%\firstname{Marcelo} \lastname{Disconzi}\inst{2}\fnsep\thanks{\email{marcelo.disconzi@vanderbilt.edu}} 
%
%\and 
%
\firstname{Jorge} \lastname{Noronha}\inst{1}\fnsep\thanks{\email{jn0508@illinois.edu}}}
%%%%%%% NAMES OF INSTITUTIONS %%%%%%%%%%
\institute{Illinois Center for Advanced Studies of the Universe, Department of Physics, 
University of Illinois Urbana-Champaign, Urbana, IL 61801, USA 
\and 
Department of Mathematics, Vanderbilt University, Nashville, TN, USA 
\and 
Escola de Ci\^{e}ncias e Tecnologia, Universidade Federal do Rio Grande do Norte, RN, 59072-970, Natal, Brazil 
\and 
Department of Physics and Astronomy, University of Florence, Via G. Sansone 1, 50019 Sesto Fiorentino, Italy}
%\date{\today}
%\pagestyle{empty}
%\preprint{CERN-TH-2023-215}
\abstract{We present the first set of fully-nonlinear, necessary and sufficient conditions guaranteeing causal evolution of the initial data for the Israel-Stewart hydrodynamic equations with shear and bulk viscosity coupled to a nonzero baryon current. These constraints not only provide nonlinear causality: they also (a) guarantee the existence of a locally well-posed evolution of the initial data (they enforce strong hyperbolicity) when excluding the endpoints of the bounds, (b) arise from purely algebraic constraints that make no underlying symmetry assumptions on the degrees of freedom and (c) propagate the relevant symmetries of the degrees of freedom over the entire evolution of the problem. Our work enforces a mathematically rigorous foundation for future studies of viscous relativistic hydrodynamics with baryon-rich matter including neutron star mergers and heavy-ion collisions.}

\maketitle

\section{Introduction}

Relativistic hydrodynamics is an effective description of a vast range of out-of-equilibrium phenomena spanning sub-nuclear length-scales in relativistic heavy-ions to neutron star mergers and accretion disks around supermassive black holes. Even so, a closer look at these constitutive equations of motion from a mathematical perspective reveals coupled systems of quasilinear partial differential equations with highly non-linear coefficients of the dynamic degrees of freedom. One such class of theories known as Israel-Stewart (IS) theory \cite{ISRAEL1976310, ISRAEL1979341} was initially proposed in the late 1970's to resolve the known acausality \cite{Hiscock_Lindblom_acausality_1987} and instability \cite{Hiscock_Lindblom_instability_1985} of the first-order viscous theories of Eckart \cite{EckartViscous} and Landau \cite{LandauLifshitzFluids} by expanding the entropy current up to \emph{second-order} in dissipative currents instead of truncating at linear order. Though, to this day, the entire region of nonlinear causality for IS-theory remains unknown, even though \emph{linear} causality and stability were established in the 1980's \cite{Hiscock_Lindblom_stability_1983, Hiscock_Lindblom_pathologies_1988}. We remark that nonlinear analyses are needed as they are able to constrain the viscous degrees of freedom directly, in contrast to linear analyses. Exact nonlinear constraints for causality \cite{ChoquetBruhatGRBook} and strong hyperbolicity (local well-posedness) \cite{ReulaStrongHyperbolic} have been derived for IS theory with only bulk\footnote{Symmetric hyperbolicity was actually shown in this work, which is a more stringent condition.} \cite{Bemfica:2019cop} and only shear when coupled to magnetic fields in the large-field limit \cite{Cordeiro:2023ljz}. Nonlinear causality constraints were recently derived for IS-theory with number diffusion as well \cite{cordeiro2025nonlinearcausalityisraelstewarttheory}. An even more general extension including both bulk and shear viscosity in the zero baryon regime for IS-like equations from kinetic theory\footnote{In fact, we allow the transport coefficients to depend on viscous fluxes e.g. $\zeta = \zeta(\varepsilon,n,\Pi,\pi_{\alpha\beta}\pi^{\alpha\beta})$, making the class of theories we consider significantly more general.} \cite{Betz:2008me} found two separate sets of conditions, with one set necessary, and the other set sufficient for nonlinear causality \cite{Bemfica:2020xym}.

In this work \cite{cordeiro2025nonlinearcausalityisraelstewarttheory}, we extend the results of \cite{Bemfica:2020xym} to allow for a non-zero baryon current and establish a set of algebraic constraints that are \emph{simultaneously necessary and sufficient} for nonlinear causality. These conditions eliminate ambiguous regions where causality is contested. In addition, we also provide a set of sufficient conditions for strong hyperbolicity. Both constraints are algebraic and make no assumptions on the spacetime metric or equation of state.
\emph{Notation:} We use a mostly-plus metric signature and natural units $c=\hbar=k_B = 1$ where $k_B$ is the Boltzmann constant.

\section{Equations of Motion}

We consider a system described by a single fluid of baryons with number density $n$ and average fluid four-velocity $u^\mu$ with normalization $u^\alpha u_\alpha = -1$, along with bulk viscous pressure $\Pi$ and traceless and symmetric shear stress $\pi^{\mu\nu}$ ($u_\alpha\pi^{\alpha\mu} = 0^\mu$). We assume a conserved energy-momentum tensor $T^{\mu\nu} = \varepsilon u^\mu u^\nu + (P+\Pi)\Delta^{\mu\nu} + \pi^{\mu\nu}$ and baryon current $J^\mu = nu^\mu$ where we ascribe $\varepsilon$, $P$ and $\Delta_{\mu\nu} = g_{\mu\nu} + u_\mu u_\nu$ to the energy density, equilibrium pressure and projection tensor orthogonal to $u^\mu$, respectively. 
%\bml
%\label{eq:conservationlaw}
%\bea
%\label{eq:energy-nodiffusion}
%0 &=& u^\alpha\nabla_\alpha \varepsilon + E\nabla_\alpha u^\alpha + \pi^{\alpha\beta}\nabla_\beta u_\alpha,\\
%\label{eq:momentum-nodiffusion}
%0^\mu &=& Eu^\alpha\nabla_\alpha u^\mu + \Delta^{\mu\alpha}\nabla_\alpha(P+\Pi) + \nabla_\alpha\pi^{\mu\alpha} - u^\mu \pi^{\alpha\beta}\nabla_\beta u_\alpha,\\
%\label{eq:number-nodiffusion}
%0 &=& u^\alpha\nabla_\alpha n + n\nabla_\alpha u^\alpha,
%\eea
%\eml
The IS-like equations from kinetic theory are provided in \cite{Betz:2008me} for bulk and shear stress as 
\bml
\label{eq:supplementalIS}
\bea
\label{eq:bulk-nodiffusion}
\tau_\Pi u^\alpha\nabla_\alpha\Pi + \Pi &=& -\zeta\nabla_\alpha u^\alpha -\delta_{\Pi\Pi}\Pi\nabla_\alpha u^\alpha - \lambda_{\Pi\pi}\pi^{\alpha\beta}\sigma_{\alpha\beta},\\
%%%
\label{eq:shear-nodiffusion}
\tau_\pi\tensor{\Delta}{^{\mu\nu}_{\beta\gamma}}u^\alpha\nabla_\alpha\pi^{\beta\gamma} + \pi^{\mu\nu} &=& -2\eta\sigma^{\mu\nu} - \delta_{\pi\pi}\pi^{\mu\nu}\nabla_\alpha u^\alpha -\tau_{\pi\pi}\pi^{\alpha\langle\mu}\tensor{\sigma}{^{\nu\rangle}_\alpha}-\lambda_{\pi\Pi}\Pi\sigma^{\mu\nu}.
\eea
\eml
Here, $\sigma^{\mu\nu} = \tensor{\Delta}{^{\mu\nu}_{\alpha\beta}}\nabla^\alpha u^\beta$ is the shear tensor in terms of the traceless symmetric projection tensor orthogonal to $u^\mu$ expressed as $\tensor{\Delta}{^{\mu\nu}_{\rho\sigma}} = \frac{1}{2}\left(\tensor{\Delta}{^\mu_\rho}\tensor{\Delta}{^\mu_\sigma} + \tensor{\Delta}{^\mu_\sigma}\tensor{\Delta}{^\mu_\rho}\right) - \frac{1}{3}\Delta^{\mu\nu}\Delta_{\rho\sigma}$. The transport coefficients $\{\zeta, \eta\}$ govern first-order deviations from equilibrium, whereas $\{\delta_{\Pi\Pi}, \lambda_{\Pi\pi}, \delta_{\pi\pi}, \tau_{\pi\pi}, \lambda_{\pi\Pi}\}$ are the second-order contributions. Together with conservation of energy $u_\alpha\nabla_\beta T^{\alpha\beta} = 0$, momentum $\tensor{\Delta}{^\mu_\alpha}\nabla_\beta T^{\alpha\beta} = 0$ and baryon number $\nabla_\alpha J^\alpha = 0$, Eq.~\eqref{eq:supplementalIS} defines our equations of motion.

\section{Nonlinear Causality and Strong Hyperbolicity}

The IS equations in Eq.~\eqref{eq:supplementalIS} and conservation laws may be uploaded into the matrix form
\be
\label{eq:FirstOrderQPDE}
(\mathbb{A}^\alpha\partial_\alpha + \mathbb{B})\Umb = \boldsymbol{0},
\ee
where $\Umb = (u^\nu,\varepsilon,n,\Pi,\pi^{\nu 0},\pi^{\nu 1},\pi^{\nu 2},\pi^{\nu 3})^T\in\mathbb{R}^{23}$ is the column vector of dynamic degrees of freedom for the system and $^T$ denotes transposition. %For convenience, any fixed raised indices will denote representative column vector elements, whereas lowered indices will correspond to row vectors.
Here, $\mathbb{A}^\mu \equiv \mathbb{A}^\mu(\Umb)$ are the $23\times 23$ matrices containing the \emph{coefficients} of the highest (first) order derivative terms but not any coordinate derivatives themselves, and $\mathbb{B}$ is a $23\times 23$ matrix containing all zeroth-order terms in coordinate derivatives. In this sense, the system is \emph{quasilinear}. Thus, one can consider the characteristic vector $\phi_\mu = \partial_\mu\Phi(x)$ as the gradient of some characteristic hypersurface $\Phi$ of the system and proceed by the method of characteristics, defined by the characteristic equation $\det(\mathbb{A}^\alpha\phi_\alpha) = 0$ \cite{Courant_and_Hilbert_book_2}. Let $x = u^\alpha\phi_\alpha$ and $v^\mu = \Delta^{\mu\alpha}\phi_\alpha$. 
We express the determinant as
\bml
\bea
\label{eq:nodiffusiondet}
\frac{\det(\mathbb{A}^\alpha\phi_\alpha)}{v^{23}\prod\limits_{k=1}^3(E + \Lambda_k)} &=& E\tau_\Pi\tau_\pi^{16}\hat{x}^{15}\left(\hat{x}^2 -\frac{\frac{1}{2} (2\eta + \lambda_{\pi\Pi}\Pi)+\frac{1}{4} \tau_{\pi\pi}\Lambda_{\hat{v}^2}}{\tau_\pi E}\right)\mathcal{P}_3(\hat{x}^2,\hat{v}_a^2),\\
\mathcal{P}_3(\hat{x}^2,\hat{v}_a^2) &=& \hat{x}^6 - \mathcal{A}_2(\hat{v}_a^2)\hat{x}^4 + \mathcal{A}_1(\hat{v}_a^2)\hat{x}^2 -\mathcal{A}_0(\hat{v}_a^2),
\eea
\eml
where we have let $v^A = (0,v^a)$ be the components of $v^\mu$ in the basis of shear-stress eigenvectors (which exists since $\pi^{\mu\nu}$ is symmetric), $\Lambda_{\hat{v}^2} \equiv \sum_{a = 1}^3 \Lambda_a \hat{v}_a^2$ and $E = \varepsilon + P + \Pi$ for convenience. Furthermore, we write the ``normalized" characteristics as $\hat{x}^2 \equiv x^2/v^2$ and $\hat{v}_A^2 \equiv v_A^2/v^2$. The coefficients of the cubic polynomial $\mathcal{P}_3$, given by $\mathcal{A}_0$, $\mathcal{A}_1$, $\mathcal{A}_2$ are highly complicated nonlinear functions of the dynamic variables in $\Umb$ and angles in $\hat{v}_A^2$. Their exact form is suppressed here for brevity.

Formally, a system of the form $(\mathbb{A}^\alpha\partial_\alpha + \mathbb{B})\Umb = 0$ is \textit{causal} if, and only if (CI) the roots of the characteristic equation $\det(\mathbb{A}^\alpha\phi_\alpha) = 0$, given by $\phi_0 \equiv \phi_0(\phi_i)$ are real, where $\phi^\mu \equiv \nabla^\mu\Phi$, and $\{\Phi(x) = 0\}$ are the characteristic hypersurfaces and (CII) $\phi^\mu$ is non-timelike, i.e. $\phi^\alpha\phi_\alpha \geq 0$ \cite{ChoquetBruhatGRBook}. Put together, condition (CI) states that the characteristic speeds must be real and exist, and condition (CII) enforces that these speeds are causal. We state our final result as
\begin{theorem}\label{thm:causalitynodiffusion}
Let $\tau_\Pi,\tau_\pi\neq 0$, $E + \Lambda_A \neq 0$ for $A = 0,1,2,3$, and $\Delta(\mathcal{P}_3;\chi^2,\kappa^2)$ be the discriminant of $\mathcal{P}_3$. If the following conditions hold simultaneously $\forall\chi^2,\kappa^2\in[0,1]$:
\bml
\bea
\label{eq:C1}
0&\leq &\Delta(\mathcal{P}_3;\chi^2,\kappa^2),\\
%%%
\label{eq:C2}
0 &\leq & \frac{\mathcal{A}_2(\hat{v}_a^2)}{3} \leq 1,\quad 0\leq \mathcal{A}_1(\hat{v}_a^2),\quad 0\leq \mathcal{A}_0(\hat{v}_a^2)\\
%%%
\label{eq:C3}
0 &\leq & 1-\mathcal{A}_2(\hat{v}_a^2) + \mathcal{A}_1(\hat{v}_a^2) - \mathcal{A}_0(\hat{v}_a^2) ,\\
%%%
\label{eq:C4}
0 &\leq & 1-\frac{2}{3}\mathcal{A}_2(\hat{v}_a^2) + \frac{1}{3}\mathcal{A}_1(\hat{v}_a^2) ,\\
%%%
\label{eq:C5}
0 &\leq & \frac{\frac{1}{2} (2\eta + \lambda_{\pi\Pi}\Pi)+\frac{1}{4} \tau_{\pi\pi}\Lambda_{\hat{v}^2}}{\tau_\pi E} \leq 1.
\eea
\eml
Then the solutions of Eq.~\eqref{eq:FirstOrderQPDE} propagate causally over their time of existence if and only if Eq.~\eqref{eq:C1}--\eqref{eq:C5} hold.
\begin{proof}
The proof will be provided shortly in a future publication.
\end{proof}
\end{theorem}
For a system of quasilinear PDEs, strong hyperbolicity \cite{ReulaStrongHyperbolic} is a desirable trait that guarantees the existence and uniqueness of the Cauchy problem \cite{ChoquetBruhatGRBook} over some finite time of existence (i.e. the problem is locally well-posed). 
%More precisely, we say a quasilinear system $(\mathbb{A}^\alpha\partial_\alpha + \mathbb{B})\Umb = \boldsymbol{0}$ is \emph{strongly hyperbolic} if, given some time-like vector $\xi^\mu$ (HI) $\det(\mathbb A^\alpha \xi_\alpha) \not= 0$, and (HII) for any space-like vector $\zeta^\mu$, the solutions of the eigenvalue equation $(\Lambda \xi_\alpha + \zeta_\alpha)\mathbb A^\alpha \mathbf r = \mathbf 0$ exist for $\Lambda\in\mathbb{R}$ and the right eigenvectors $\mathbf{r}$ span a complete basis. The first condition requires the causal propagation of solutions (nonsingular principal part for timelike normal vectors), whereas the second condition is an extra, useful condition providing a set of complete basis of eigenvectors. 
If instead, one replaces all inclusive lower bounds `$0\leq\cdots$' in Eq.~\eqref{eq:C1}--\eqref{eq:C2} and \eqref{eq:C5} with the strict conditions `$0 <\cdots$', then Eq.~\eqref{eq:C1}--\eqref{eq:C5} provide a set of sufficient conditions for strong hyperbolicity. We present the proof for this statement in a forthcoming work.

\section{Conclusion}

Our analysis presents the first general nonlinear bounds on causality for a very general class of IS-like theories with bulk and shear contributions. This class allows for (i) all transport coefficients to depend directly on $\Pi$ and $\pi^{\mu\nu}$ themselves, (ii) extra terms obtained through kinetic theory, and (iii) matter in which baryon number is non-zero (e.g. neutron star and accretion physics). In particular, we make no assumptions about the spacetime metric or equation of state. Thus, our results encompass traditional IS \emph{in addition} to a significantly more diverse class of both theories and physical systems. We also show that a large subset of the causality constraints imply the existence and uniqueness of solutions over some finite timescale for a very general class of functions (strong hyperbolicity). These results make significant headway in understanding the nonlinear validity of IS theories. 

\noindent
\emph{Acknowledgements.}
ES has received funding from the European Union’s Horizon Europe research and innovation program under the Marie Sk\l odowska-Curie grant agreement No. 101109747. JN and IC are partly supported by the U.S. Department of Energy, Office of Science, Office for Nuclear Physics under Award No. DE-SC0023861. This work was done while the author FSB was a Research
Assistant Professor at Vanderbilt University.
%This research was partly supported by the National Science Foundation under Grant No. NSF PHY-1748958 and NSF PHYS-2316630. 
%Any opinions, findings, and conclusions or recommendations expressed in this material are those of the author(s) and do not necessarily reflect the views of the National Science Foundation.

\def\cprime{$'$}

\end{document}